 \documentclass[preprint2]{aastex}
\usepackage{txfonts}

\slugcomment{Not to appear in Nonlearned J., 45.}

\def\dens{g.cm$^{-3}$}

\shorttitle{Constraining deflagration models of Type Ia supernovae....}

\shortauthors{Garc\'\i a-Senz, Bravo, Cabez\'on and Woosley}

\begin{document}


\title{Constraining deflagration models of Type Ia supernovae 
              through intermediate-mass elements}


\author{D. Garc\'\i a-Senz\altaffilmark{1,2}, 
E. Bravo\altaffilmark{1,2}, R.M. Cabez\'on\altaffilmark{1}
   and S.E. Woosley\altaffilmark{3}}
\altaffiltext{1}{Departament de F\'\i sica i Enginyeria Nuclear, UPC, 
Jordi Girona 3, M\`odul B5, 08034 Barcelona, Spain; domingo.garcia@upc.edu; eduardo.bravo@upc.edu; ruben.cabezon@upc.edu} 

\altaffiltext{2}{Institut d'Estudis Espacials de Catalunya, Barcelona, Spain}
\altaffiltext{3}{Department of Astronomy and Astrophysics, 477 Clark Kerr Hall, University of California, Santa Cruz, CA 95064; woosley@ucolick.org}



\begin{abstract}
The physical structure of a nuclear flame is a basic ingredient of the
theory of Type Ia supernovae (SNIa). Assuming an exponential density 
reduction with 
several characteristic times we have followed the evolution 
of a planar nuclear flame in an expanding background from an initial density  
$6.6\times 10^7$~\dens down to $2\times 10^6$~\dens. 
The total amount of synthesized intermediate-mass elements (IME), from silicon
to calcium, was monitored during the calculation. 
We have made use of the computed mass fractions, X$_{IME}$, of these 
elements to give an estimation of the  total amount 
of IME synthesized during the deflagration of a massive white dwarf. Using 
X$_{IME}$ and adopting  
the usual hypothesis that turbulence decouples the effective burning 
velocity from the laminar flame speed, so that the relevant flame speed is 
actually the turbulent speed on the integral length-scale, 
we have built a simple 
geometrical approach to model the region where
IME are thought to be produced.  
It turns out that a healthy  
production of IME involves the combination of not too short 
expansion times, $\tau_c\ge 0.2$~s, and high turbulent intensities. According 
to our results it could be difficult to produce much more than 0.2 M$_\sun$~of 
intermediate-mass elements within the deflagrative paradigma. The calculations 
also suggest that the mass of IME scales with the mass of Fe-peak elements,  
making it difficult to conciliate energetic explosions with low ejected nickel 
masses,  
as in the well observed  SN1991bg or in SN1998de. Thus 
a large production of Si-peak elements, especially in combination with a low or a moderate production of iron, could be better addressed by either   
the delayed detonation route in standard Chandrasekhar-mass models or, perhaps, 
by the 
off-center helium detonation in the sub Chandrasekhar-mass scenario.

\end{abstract}




\section{Introduction}

A distinctive feature of SNIa spectra near maximum light is the presence of 
intermediate-mass elements from Si to Ca moving at velocities
 10000-16000 km.s$^{-1}$. In particular the presence or not of the SiII 
absorption feature at $6150 \AA$~ 
is basic to  
identify a Type I supernovae as SNIa. According to observations 
about 0.2-0.4 M$_\sun$ of IME have to be synthesized during the explosion. An 
amount of IME larger than 0.2M$_\sun$~is also needed to explain the observed 
abundances of these elements in the solar neighborhood  
\cite{iwa99}.  
Therefore, any attempt to build a satisfactory theoretical model of 
 these explosions has to give the right amount of intermediate-mass elements.  

In this respect the underproduction of intermediate-mass elements 
found in current pure deflagration 
models of Type Ia supernovae is a long standing problem of theoretical 
modelization of these explosions. 
On fundamental basis we know that these 
elements have 
to be synthesized below densities $\rho\simeq 5\times 10^7$~\dens in an 
expanding medium. Nevertheless the resolution achieved by the best
 multidimensional models carried out to date is far from adequate to solve 
unambigously this problem. In this work we have taken another route to 
address this issue. Instead of performing a multidimensional calculation 
incorporating the unresolved scales as a subgrid we have simulated the 
evolution of an expanding  nuclear flame in planar aproximation 
with  very 
good resolution. We have then used the results to put limits to the amount of IME 
synthesized in a deflagration. The usefulness of our approximation 
requires that the synthesis of elements within the mass
 range $28\leq$A$\leq 40$~
takes place, for the most part, during the flamelet regime. In such a regime  
the flame, although highly wrinkled and distorted, still preserves its 
identity. Unfortunately the precise value of the minimum density at which the 
flamelet hypothesis still holds is not known. There are indications that 
transition from the flamelet to the distributive regime in the combustion 
diagram could begin 
around $2\times 10^7$~\dens \citep{nie99, zin05}, and we can take it as a reference. 
 
Keeping that constraint in mind we have proceeded  
in the following way. First, the detailed structure and evolution of a 
nuclear flame propagating through an expanding background has been solved for 
a reasonable set of characteristic expansion times and the  
nucleosynthetic yields calculated using a simplified $\alpha$-network of 14 
nuclei, $\S\ 2$. With that information we provide in $\S\ 3$~ a rough 
calculation of the total
 amount of IME 
synthesized in a typical deflagration of a white dwarf as a 
function of the effective surface of heat exchange between ashes and fuel at 
the densities of interest, the turbulence intensity and the characteristic expansion time. According to our results  
a subsonic deflagration is able to produce more than 0.2M$_\sun$~of 
IME although only in a limited region of parameter space. This point is discussed in the chapter devoted to the
 conclusions, $\S\ 4$, as well as possibles issues and limitations of our work. 

\section{Production of Intermediate-mass elements in an expanding laminar 
        flame} 

From the nucleosynthetic point of view IME are the main product of  
incomplete Si burning at 
$T_9\leq 5$~ followed by $\alpha$-rich 
freeze-out until the quenching of nuclear reactions. During the explosion 
of a massive white dwarf such combustion regime is attained when the density ahead the flame drops below 
$\simeq 5\times 10^7$~g.cm$^{-3}$~in a medium which is rapidly expanding. As density 
decreases, the laminar velocity of the flame goes down,  
its width enlarges and the nucleosynthetic products are enriched first in IME 
 and later in $^{24}$Mg~and $^{16}$O. Finally  
the flame becomes  
a thermal wave and no more nuclear combustion takes place: {\sl the flame dies}.
Even though the thickness of the zone where IME are synthesized in the planar 
case is already 
really  narrow, around half kilometer (see below),  
the surface of the flame in a real situation is so distorted by
 hydrodynamical instabilities that the 
effective volume might be enough to account for the observed amount of IME. 
One important consequence is that, once developed, the hydrodynamical 
instabilities make the effective 
burning velocity independent of the laminar velocity of the flame (\S\ 3). 
   
We have devised a numerical experiment to investigate whether or
not the adequate amount of IME can be produced by the above  
 mechanism.
First of all we have made an estimation of the total amount of chemical species generated by a planar flame moving through a rapidly expanding media. Although 
in the end the effective turbulent burning velocity will become independent 
of the laminar value such calculation is worthwhile because: 1) it provides 
the mass fraction of the isotopes of interest to be used in the next section 
when turbulence effects are taken into account, 2) It shows the futility of 
trying the synthesis of IME in a real explosion when the flame only moves at 
its laminar speed, and  
3) knowing the physics of an expanding planar nuclear flame is an 
interesting topic by itself as it represents a natural extension of the 
work by Timmes \& Woosley (1992) (TW afterwards). The 
ouptut of the expanding planar flame calculation  gives a fundamental quantity: 
$\Delta m_{IME}$, the 
mass of IME synthesized in a column with 1 cm$^2$~section. For a 
given amount of burnt material $\Delta m_{burnt}$~we compute the 
mass fraction of IME, X$_{IME}$~synthesized below a fixed starting 
density $\rho_0$~for different expansion rates. Later
on (\S\ 3)~ 
we make use of X$_{IME}$~to get an estimation of the total amount of 
IME ejected during a typical Type Ia supernova event. The  
quantity $\Delta m_{IME}$~was calculated by following the propagation of the 
 nuclear 
flame in planar geometry as explained in 
Cabez\'on et al. \citep{cab04}. 
Basically the integration scheme uses an   
 $\alpha$-network with 14 nuclei and {\sl implicit thermal coupling}
 with the 
energy equation to
follow the combustion under isobaric conditions. The resulting method is 
robust and stable, allowing to properly handle the rear tail of the flame 
where most of the elements of interest were synthesized. In addition  
an  exponential reduction in density ahead the flame,  
with a characteristic time $\tau_c$, is imposed. The  
resulting set of equations are:

\begin{equation}
\frac{dY_i}{dt}=\sum_{k,l}r_{kl}Y_k Y_l-\sum_j r_{ij} Y_i Y_j+
   \sum_m \lambda_m Y_m-\lambda_i Y_i
\end{equation}

\begin{equation}
\frac{dU}{dt}=\frac{1}{\rho}\frac{\partial}{\partial x}
\left(\sigma\frac{\partial T}{\partial x}\right)+\frac{P}{\rho^2}\frac{d\rho}{dt}
+\dot\epsilon_{nuc}
\end{equation}

\begin{equation}
\rho (x>>x_{flame-front})=\rho_0~\exp[-t/\tau_c]
\end{equation}

\begin{equation}
P(t)=P_{ext}(\rho, T)\qquad\qquad
\end{equation}

\noindent
 where $r_{ij}$~stands for particle reaction rates,  
$\lambda_i$~ is for photodisintegrations and the rest of the symbols have 
their usual meaning.
The thermal conductivity $\sigma$~was taken from  Appendix 1 in  
 Khokhlov et al. (1997) and we have taken an uniform initial density of the 
fuel 
 $\rho_0=6.6\times 10^7$~\dens. The precise value of $\rho_0$~is irrelevant 
for the results as long as it is larger than the density at which 
IME begin to be produced. 
At each step we enforce the temperature of the unburnt material, 
located far 
enough from the flame front, to evolve adiabatically. 
 The instantaneous 
value of T and $\rho$ far ahead the flame sets the value of P$_{ext}$~in 
Equation (4). Across the flame the evolution is governed by  
heat difussion and binary $^{12}C+^{12}C$~reaction, whereas in the burned 
zone far from 
the flame front the incinerated material relaxes to either Fe-peak elements 
or to IME if density becomes low enough. The integration scheme is a 
standard lagrangian code with enough mass-shells to resolve the structure of 
the flame. At the beginning of the calculation a small group of shells were 
put into nuclear statistical equilibrium at constant pressure. Very soon a 
steady nuclear flame developed and propagated through the cold fuel (composed 
by equal parts of carbon and oxygen). As the flame progressed we kept a constant
number of mass-shells ahead the flame (around 100) in order to save
computer time. 
Finally the calculation was stopped when the 
temperature in the burned zone dropped below two billion degrees and all
nuclear reactions virtually quenched. At the end of the calculation there 
were typically between $2\times 10^4$~to $10^5$~shells in combustion, 
depending on the characteristic expansion time $\tau_c$. It was necessary to 
calculate several millions of time steps in order to follow the whole process. 

In Figure 1 it is shown a snapshot of the  mass fraction profile at
 t=0.1523 s, which is the time at which total IME abundance freezes in  
our reference model with $\tau_c=0.1$~s. As we 
can see $^{28}$Si as well as other IME with higher atomic mass were synthesized
in a tiny layer, lesser than 0.3 km,  
which is more than one order of magnitude below the resolution of 
the best three-dimensional simulations carried out so far. Moreover because 
during the explosion the flame is shattered by turbulence that length 
represents an upper limit of the size of pockets of ashes made of 
elements belonging to the silicon family.  
Elements lighter than silicon such as magnesium and neon 
were produced at lower density in a very narrow region behind the flame edge. 
In Figure 2 we depict the evolution of the 
total synthesized mass of elements between $^{28}$Si~to $^{40}$Ca for
$\tau_c=0.05,~0.1,~ 0.2~ \mathrm{and}~ 0.3$~s 
as a function of the fuel density. From these figures it is 
clear that for plausible expansion times the IME are synthesized in 
the density range $2-5\times 10^7$~\dens, as it was recognized many 
years ago. In particular, we see that  
95\% of IME were already synthesized at densities higher than $2\times 10^7$~
\dens. In fact, more than a half of their final mass was produced 
when the density of the fuel was still above $4\times 10^7$~\dens. This is an 
interesting feature because it suggests that important nuclei for 
SNIa spectroscopy, such as 
silicon and calcium, are probably synthesized during the so-called flamelet 
regime of 
turbulent combustion, much more easier to handle than the distributive regime. 
Below $2.5\times 10^7$~\dens we found an 
increase of 
direct carbon and oxygen combustion products, neon and magnesium in our network, in a region of size $\simeq 0.05$~km. 
However 
in a real physical situation the final abundance of elements below silicon
 might be altered 
if combustion enters the distributive regime. 
The velocities of the flame, fitted using the total amount of burnt fuel (see Eq.7 below), are in reasonable  
agreement with those obtained by TW 
 at different densities. We obtain slightly lower values because the limited nuclear 
network used to carry out the calculations. For example, we obtained v=0.14, 
0.23, and 
0.33 km.s$^{-1}$~at $\rho_7$=2, 3 and 4 respectively, to be compared  
 to those calculated by TW,  
v$_{TW}$=0.19, 0.32, 0.46 km.s$^{-1}$~at the same densities. 
The instantaneous width of the flame was also consistent with TW calculation. 
The 
detailed final yields of the 14 nuclei for several characteristic 
expansion times are shown in Table 1. As we can see, there is a monotonic 
increase of  
the final mass of the most abundant species with the expansion time 
$\tau_c$. For $\tau_c=0.05$~s we got $2.95\times 10^{10}$~g.cm$^{-2}$~ of
 silicon 
whereas for $\tau_c=0.3$~s the synthesized mass was $1.16\times 10^{11}$~g.cm$^{-2}$. 
The same trend is true for the size of the region where these elements 
are significatively produced: $\simeq 0.15$~km ($\tau_c=0.05$~s) to 
$\simeq 0.75$~km ($\tau_c=0.3$~s). Even in the last case the width of the 
IME layer is much smaller than the size of the burned region,
typically 2000 km.  
Thus in the lagrangian one-dimensional picture it can be regarded as a very 
thin layer
 on top of a volume 
made basically of Fe-peak elements.  

The mass of fuel consumed during the propagation of an expanding 
planar flame starting at an 
initial density $\rho_0$~can be found analytically because the density 
dependence 
of the laminar velocity is known (TW):

\begin{equation}
v=v_0\left(\frac{\rho}{\rho_0}\right)^n
\end{equation}

From our results $v_0\simeq 0.6$~km.s$^{-1}$~at $\rho_0=6.6\times 10^7$~\dens, 
and $n\simeq 1.2$. Using Equation 3 the mass of burned fuel per unit surface in 
an elapsed time $\Delta t$~is:

\begin{eqnarray}
\Delta m_{burnt}=\rho_0~v_0\int_{0}^{\Delta t}\exp{\left[-\frac{(1+n)t}{\tau_c}
\right]}~dt=\\ \nonumber
\rho_0~v_0~\frac{\tau_c}{(1+n)}\left[1-\exp{\left(-\frac{(1+n)\Delta t}{\tau_c}\right)}\right]
\end{eqnarray}

\noindent
Taking $\Delta t\rightarrow\infty$~in Equation 6 gives a good approximation to
the the total burnt mass per unit surface:

\begin{equation}
\Delta m_{burnt}=\rho_0 v_0\frac{\tau_c}{(1+n)}
\end{equation}

\noindent
Hence the amount of consumed fuel below the 
starting density $\rho_0$~in the expanding background is 
simply proportional to the product of the laminar velocity and the 
characteristic expansion time. The laminar 
velocity of the flame is rather well known within, at most, a factor two
 uncertainity. The 
 value of the characteristic expansion time $\tau_c$~is probably 
dependent of the particular explosion dynamics. A rough lower limit for 
$\tau_c$~is 
the characteristic hydrodynamical time scale $\tau_h=446/\sqrt{\rho}=0.08$~s at $\rho=3\times 10^7$~\dens which matches well also the behaviour of a particular 
deflagrative one-dimensional model \citep{bra96} as shown in Figure 3. 
A reasonable range for $\tau_c$~is    
$0.05 s\le\tau_c\le 0.3 s$~once the flame has reached densities below $6\times 
10^7$~\dens,   
although  higher values, nearing 0.5 seconds, can not be totally  
discarded.  
In Table 2 there is shown the total amount of burned mass as well as the
 masses  
of IME and nickel synthesized below $\rho_0$~as a function of $\tau_c$. As we 
can see the burnt mass scales linearly with $\tau_c$, as suggested by Equation 
7. 
However,  the   
amount of synthesized IME is not proportional to the characteristic expansion 
time because a longer value of $\tau_c$~leads to the 
formation of more iron as the nuclear system has more room to relax. A simple  
fitting formulae to the values of $\Delta m_{IME}$~given in Table 2 that can be
 also used to extrapolate $\Delta m_{IME}$~beyond 0.3 s, until
 $\tau_c\simeq 0.5$~s if necessary, is: 

\begin{equation}
\Delta m_{IME}=6.5\times 10^{11}~\tau_c^{\frac{3}{4}}~{\mathrm g.cm^{-2}}
\end{equation}

As mentioned before, the mass of IME given by Equation 8 is independent of the 
particular value
 chosen for $\rho_0$~in Equation 3 provided it is high enough to resolve the 
region where intermediate-mass elements are synthesized. However their   
mass fraction, X$_{IME}$, within the burnt material is a function of the 
starting density $\rho_0$~in the exponential expansion. For 
$\rho_0=6.6\times 10^7$~\dens~it can be fitted using:

\begin{equation}
X_{IME}=0.35~\tau_c^{-\frac{1}{4}}
\end{equation}

Another remarkable feature shown in Table 2 is that the production 
of $^{56}$Ni is no longer negligible, 
especially 
for $\tau_c\ge 0.2$~s. At the longest expansion time considered here, $\tau_c=0.3$~s, the amount of $^{56}$Ni is practically equal to that of IME.

\section{Consequences for Type Ia Supernova models} 

A quantitative estimation of the total amount of IME ejected in a SNIa 
explosion may be done provided the effective flame surface at densities   
$\rho\simeq 2-5\times 
10^7$~\dens is somehow known. However, this is a difficult task that will 
have to  
be addressed in the future through high resolution multidimensional
 simulations. 
For the time being we give only a qualitative discussion about this point 
taking advantage of the self-similar behaviour
 of the main  physical
 agents  responsible 
for the corrugation of the flame. There has been shown  
\citep{kho95, nie97}, that a large increase
in flame surface can only be attributed to two physical mechanisms: the 
Rayleigh-Taylor (RT) 
instability and turbulence. Fortunately, it turns out that both are 
self-similar phenomena that can be approximately described using  
three parameters: the 
maximum and minimum scale-lengths, $\lambda_{max}, \lambda_{min}$~and the 
average fractal dimension D' of the corrugated flame surface. In this approximation 
the incinerated mass is given by: 

\begin{equation}
M_{burnt}=4\pi \bar{r}^2\left<\frac{\lambda_{max}}{\lambda_{min}}\right>^{D'-2}\int_0^{\Delta t}\rho~v_{flame}~dt
\end{equation}

\noindent
where $\bar{r}$~is the average radius of the flame and $v_{flame}$~stems for 
its laminar velocity. Even though in the supernova the nuclear flame becomes 
highly distorted and shattered by turbulence and hydrodynamical instabilities 
the amout of fuel consumed per unit area given by the integral part of 
Equation 10 should not be very different from $\Delta m_{burnt}$~given by 
Equation 7 corresponding to the planar case analyzed in the 
previous section. However many experiments in combustion chambers as well as numerical simulations dealing with turbulent combustion suggest that 
$\lambda_{min}$~is no longer an independent quantity \citep{pet89}. Its local
 value self-adapts so that the effective burning rate is set by the integral 
length $\lambda_{max}$~and the characteristic velocity, V, at the  
integral length-scale. Such interesting behaviour relies in the existence of 
an universal scaling law for velocity spectrum of the type:

\begin{equation}
\left(\frac{\lambda_{max}}{\lambda_{min}}\right)=
\left(\frac{V}{v_{flame}}\right)^{\frac{1}{D'-2}}
\end{equation}

\noindent
which for D'=7/3 becomes the popular Kolmogorov scaling-law for turbulence. 
Adopting the scaling relationship given by Equation 11,  both  M$_{burnt}$~
as well as  the mass of the k-isotope, M$_k$~become   
independent of $v_{flame}$. The ejected mass of the k-isotope    
 can therefore be calculated using the value of X$_k$~computed in the 
previous section:   

\begin{equation}
M_{k}=4\pi \bar{r}^2~V~\left(\int_0^{\Delta t_{k}}\rho(t)~dt\right)~X_{k}
\end{equation}

\noindent
where $\Delta t_k$~is the elapsed time until the abundance of the 
respective nuclei freezes. For example, as stated in \S\ 2 the nuclear 
combustion is unable to 
synthesize nuclei beyond magnesium once the density has gone down 
$\simeq 1.5\times 10^7$~\dens. Within our model of exponential density 
reduction this  
occurs once  $\Delta t_{IME}=1.49~\tau_c$. 
 
Admittedly, the above discussion is a little artificial, but necessary to develop 
the formalism, because finally neither v$_{flame}$~nor D' appears in 
Equation 12. 
According to that equation the progression of the flame can be envisaged as a 
spherical flame front of radius $\bar r$~moving not at the laminar 
speed but at velocity V. As V is expected to be many orders of magnitude 
larger than $v_{flame}$~the synthesized mass 
 of nickel  
as well as the released nuclear 
energy are sufficient to account for the supernova light curve and energetics. 
Nevertheless a 
variation of the geometrical approach  given by Equation 12 
could be more 
fruitful to address the particular problem of the synthesis of IME \citep{woo01}. In this regard it 
might be more physical to interpret $\bar r$~as the average radius of the 
ashes, made basically of nuclear-statistical equilibrium elements (NSEE).
The intermediate-mass elements 
are then produced in a tiny shell, almost a surface, on top of the huge 
volume defined by the underlying ashes, 
over a length 
given by the product V~$\Delta t_{IME}$. 
Unlike the synthesis of NSEE, which 
as bounded to a volume are reasonably given by Equation 12, surface 
effects are important to properly model the synthesis of IME.  
Assuming that the effective 
surface of NSEE behaves as a fractal in which those scale-lengths below 
the current thickness of the IME layer itself do not appreciably contribute 
to M$_{IME}$, we can write:   
 
\begin{equation}
M_{IME}=4\pi \bar{r}^2~\left(\frac{\bar r}{V~\Delta t_{IME}}\right)^{D-2}~V~\left(\int_0^{\Delta t_{IME}}\rho(t)~dt\right)~X_{IME}
\end{equation}

\noindent 
which either for D=2 or for (V~$\Delta t_{IME})\simeq\bar r$~reduces to 
Equation 12. 
Here, the fractal dimension $D$ is intended to describe the geometrical
properties of the NSEE external surface in between the integral length-scale and
the radius of the front, and has not to be the same as the dimension $D'$
characterizing the front at lower length-scales.
Finally, to compute M$_{IME}$~from
 Equation 13 it is more convenient to use the mass of 
the ashes made of NSEE, $M_{NSEE}$, instead their average 
radius $\bar r$. Then, solving for the integral, taking X$_{IME}$~from Equation (9) and using  
$\Delta t_{IME}=1.49~\tau_c$~the resulting 
expression is: 

\begin{equation}
M_{IME}(M_{\sun})=0.253~\left(\frac{M_{NSEE}(M_{\sun})}
{\bar\rho_7}\right)^{D/3}~
 2.43^D~V_8^{3-D}~\tau_c^{11/4-D}
\end{equation}
    
\noindent
where $\bar \rho_7$~is the average density of the ashes in
$10^7$~g$\cdot$cm$^{-3}$ and V$_8$~the 
average  
velocity of the effective flame velocity in $10^8$~cm$\cdot$s$^{-1}$.

 The main results of 
a parametric exploration of Equation 14 are depicted in Figure 4. In this 
figure there is shown the synthesized mass of elements between  silicon to 
calcium as a function of the values adopted by the set \{V$_8$, D, $\tau_c$\}.  
In computing Figure 4 we have taken a fixed mass of ashes 
$M_{NSEE}(M_{\sun})=0.8$~ 
and $\bar\rho_7=7$, a density which is  
slightly higher than the starting density during the exponential 
expansion. Such combination of M$_{NSEE}$~and $\bar\rho_7$~leads to an average 
radius of the ashes $\bar r=1.8\times 10^8$~cm, in fair agreement to 
detailed hydrodynamical calculations. For instance, model W7 \citep{nom84} gives approximately  $r_8\simeq 2$~at $\rho_7=4$~
similar as the one adopted in this work. Going to multidimensional models 
R\"opke et al. (2006) give an average radius of the burnt zone also close to $\bar r_8=2$~in 
their model c3\_3d. 
 
First of all note that the case D=2 in Figure 4 
almost always leads to an underproduction of IME. Thus it becomes  
clear why  
those one-dimensional models which assumed a spherical burning 
front moving 
at the largest turbulent speed, V, did not succeed in synthesizing large 
amounts of IME \citep{nie97}. Also, there is not a sufficient abundance of IME neither for   
$V_8<0.5$~nor for $\tau_c\le 0.05$~s.  The mass of these elements smoothly rises  
as D, V or $\tau_c$~increases, as expected. According to Equation 14 the 
most influencial parameter affecting the nucleosynthesis of silicon family 
is the fractal dimension 
of the underlying 
iron core, 
followed by the propagation velocity and the characteristic expansion time 
respectively. 

There are, however, some 
limitations  which could prevent a large production of IME during the 
explosion. 
First, the fractal dimension defining the ashes's surface is thought to be 
 closer to 
2.3 than to 2.5 because it is highly probable for turbulence to be of 
Kolmogorov type. Actually, one can expect that the flame behaves as a 
multifractal, so the effective fractal dimension would be somehow between 2.33 
(characteristic of lower scale-lengths) and D=2.5 (characteristic of large 
scale-lenghts). On the other hand the local sound speed puts an upper 
limit to the
 largest turbulent velocity V. At the densities of interest the sound speed is 
$\simeq 4-4.5\times 10^8$~cm.s$^{-1}$~thus V$_8\simeq $2 is almost a half of the 
sound speed. In that case the induction of a detonation because of the 
cumulative effect of pressure waves can not be ruled out \citep{ple04}. Maybe   
more interesting are the constraints related to the dominant combustion 
regime: 
flamelet or distributed. Taking $\lambda_{max}=\bar r=1.8\times 10^8$~cm, 
$\lambda_{min}\simeq 0.1$~cm (approximately the flame thickness at these 
densities),
 $v_{flame}\simeq 5\times 10^4$~cm.s$^{-1}$~ and D'=7/3 in Equation 11 the corresponding 
turbulent velocity is V$\simeq 6\times 10^7$~cm.s$^{-1}$. With that choice of 
parameters turbulent combustion will stay in the flamelet regime as long as the 
velocity of the largest eddy is lower than $6\times 10^7$~cm.s$^{-1}$. Were this 
the case an inspection 
of Figure 4 reveals that, even in the most favourable case with $\tau_c=0.3$~s, 
the production of IME hardly arrives
 to 0.2M$_\sun$. Considering stronger turbulent intensities would move   
the more plausible region for producing IME just to the poorly 
known distributive regime.    

The influence of the characteristic 
expansion time $\tau_c$~is not as strong as that of V$_8$~owing to the 
inverse dependence of X$_{IME}$~with $\tau_c$~(Eq. 9). Moreover, 
one-dimensional 
simulations indicate that the characteristic 
expansion time at the time combustion arrives to layers with $\rho\simeq 
5\times 10^7$~\dens are probably not much larger than 0.2 s (see for 
example Fig. 3). Even though $\tau_c$~should be relatively 
model independent, owing the strong sensitivity of the degenerate matter 
against perturbations, its precise value is set by the central density 
at the moment of the explosion and by the nuclear combustion 
dynamics itself. 

Equation 14 suggests that M$_{IME}$~and M$_{NSEE}$~are not independent 
quantities, M$_{IME}\propto M_{NSEE}^{D/3}$~. On the other hand some 
mass of Fe-peak elements are also produced simultaneously to silicon 
during the expansion. Thus, the final quantity of Fe-peak, $M_{Fe}$~
is slightly higher than $M_{NSEE}$. Assuming that after freezing is 
$X_{Fe}\simeq 1-X_{IME}$~in the IME layer 
it is easy to show that the total mass of the iron group is:

\begin{equation}
M_{Fe}(M_{\sun})=M_{NSEE}(M_{\sun})+M_{IME}(M_{\sun})~\left(\frac{\tau_c^{1/4}}{0.35}
-1\right)\qquad 0.05\le\tau_c\le 0.3
\end{equation}

A plot of $M_{IME}$~versus $M_{Fe}$~for several characteristic expansion times  at constant D=7/3 and V$_8$=1.5 is given in Figure 5. As we can see the relationship between 
$M_{IME}$~and $M_{Fe}$~is close to linear with increasing 
slope as $\tau_c$~rises. As a general rule we expect that deflagrations with 
large luminosities (i.e a large mass of  $^{56}$Ni) will also show a large 
production of IME. Nevertheless some attention has to be paid to the influence  
of the characteristic expansion time. For example, according to 
Figure 5, 0.2M$_\sun$~of IME can be synthesized in combination to 
0.7M$_\sun$~of 
Fe-peak elements at 
$\tau_c=0.3$~s or to 0.95M$_\sun$~at $\tau_c=0.1$~s. Hence explosions 
with different maximum luminosities and global energetics could lead to  
similar amounts of IME. Moreover, it is reasonable to think that 
$\tau_c$~and M$_{NSEE}$~
are inversely correlated: a large amount of M$_{NSEE}$~means a big deal   
of released nuclear energy  thus leading to shorter characteristic expansion 
times. Such correlation suggest that some self-regulating mechanism could be 
at work reducing the dispersion in the production of IME.         
 
Finally, note that our assumption that ashes are enterily composed  of NSEE has 
not to be necessarily true. Even though there is a current debate concerning 
this point, recent 
multidimensional simulations indicate that 
some unburnt carbon and oxygen could remain mixed with the  incinerated 
material, \citep{gar05, rop06}. In that case the correlation between the mass of IME and
 Fe-peak 
elements will be different to that shown in Figure 5.

\section{Conclusions}

The amount of intermediate-mass elements ejected in a SNIa explosion is an 
important diagnosing tool to discriminate between a variety of theoretical 
models attempting to explain the nature of such explosions. One of the most 
popular models is the deflagration model, rooted in the subsonic propagation
 of a nuclear flame   
through the white dwarf interior. However, with very few exceptions, neither
the one-dimensional simulations nor modern three-dimensional calculations 
of this model have been able to give the right quantity of IME. 
Hydrodynamical calculations 
always stubbornly give a too low production. This fact is currently 
attributed to the low resolution achieved in multidimensional calculations.

In the first part of this paper we have analyzed the synthesis of elements 
from helium to 
zinc during the propagation 
of a nuclear flame in an expanding background. The simulation was carried 
out in the planar aproximation, thus the resolution was enough to 
resolve the detailed structure of the flame. In this sense our work 
represents a continuation of that by Timmes \& Woosley (1992). As a 
result we were able to get an idea of the size of the region where the 
IME were 
for the most part synthesized: between 0.15 km to 0.75 km depending on the
 characteristic expansion time. This width is hardly reachable by today 
multidimensional hydrocodes. 
It was also possible to give the amount of 
elements from silicon to calcium cooked behind the flame (Table 1). This 
information can be used by SNIa modelers to improve the nucleosynthesis 
output of their hydrodynamical 
simulations. An interesting result is that IME are 
massively produced in the range  $\rho=2\times 10^7-5\times 10^7$~\dens. In fact, 
according to our 
calculations 80\% of the IME were already produced before the density declined 
below $3\times 10^7$~\dens (see Fig. 2). This could have important consequences 
because the usefulness 
of our results strongly relies in the basic hypothesis that the flame 
does not enter the distributive regime above 2-3$\times 10^7$~\dens. Finally 
useful fitting formulae giving the mass of IME synthesized during the expansion as 
well as to their mass fraction, X$_{IME}$, as a 
function of the characteristic expansion time $\tau_c$~were also given (Eqs. 8,9). 

In \S\ 3 we did use the information given in Table 1 to estimate  
 the quantity of IME ejected during the deflagration of a white dwarf taking 
advantage of the self-similar properties of the thermonuclear combustion in 
SNIa. We have taken a geometrical approach to address this problem in which 
the narrow region where the Si-peak nuclei are typically produced was treated
as a tiny shell wrapping around a deformed surface on top of a  volume 
made of Fe-peak 
elements. That approach holds as long as the ashes 
form a well defined and 
not too dispersed pack limited by an irregular surface.
Within our simple model the total amount of elements from silicon to calcium 
can be obtained from an analytical expression (Eq. 14) which uses four 
parameters: the ratio between the mass and average density of the NSE-ashes, 
the fractal dimension of their surface, the velocity of the turbulent 
combustion at the largest scales, and the characteristic expansion time. In 
consequence we have tried to delimitate the regions in parameter space able 
to give the amount of IME compatible with observations (Fig. 4).     

As a general conclusion we state the difficulty of pure deflagration models to 
produce more than 0.2M$_\sun$~of intermediate-mass elements. In this sense our 
phenomenological approach may help to understand why the underproduction 
of Si-peak elements has traditionally been a long standing problem of all 
sort of hydrodynamical simulations concerning Type Ia supernovae assuming the 
deflagrative paradigma. Still there remains a significant area in space 
parameter 
showing a rich production of IME. In particular the region defined by the 
unequalities $\tau_c\ge 0.2$~s, $D\ge 2.2$~and $V_8\ge 1$~seems to be well  
suited for that purpose. The above lower limit in the fractal dimension of the 
ahes's 
surface does not pose a problem as experiments dealing with turbulent 
combustion indicates a favoured value of D close to 2.3. 
As an aside, 1D simulations that always assume $D=2.0$ on the resolved
length-scales lead unavoidably to an underproduction of IME.
Perhaps more demanding is the limitation 
imposed by the characteristic expansion  
time. Even though $\tau_c$~should not be too different from the 
hydrodynamical time scale, $\simeq 0.1$~s at the densities of interest, 
its particular value is model dependent. Hence expansion times 
with $\tau_c\simeq 0.2$~s are probably not 
unrealistic.   
Turbulent intensities characterized by $V_8\ge 1$~are also reachable 
during the explosion because the rising velocity of the larger floating blobs 
of incinerated material is of the order, or even greater, than that velocity. 
Nevertheless, the turbulent cascade generated by these fast moving blobs could
fragment the flame in very small pieces, shorter than the flame thickness, 
invalidating our hypothesis that combustion always takes place in the 
flamelet regime.
Thus, although our main goal was to highlight the favoured regions in parameter 
space to synthesize a good deal of IME, the    
presented results also underline the potential 
importance of the distributed regime for deflagration models, if they were to  
represent the main channel to normal Type Ia supernovae.  

Even if a large mass of IME were attained in deflagration models of SNIa, our
study still suggests another potential problem: the production of IME seems to
increase monothonically with that of Fe-peak elements. Hence, it would be
seemingly difficult for deflagrations to explain peculiar low-luminosity but
energetic events like SN1991bg or SN1998de \cite{str06}.

\acknowledgements

This work was funded by Spanish DGICYT grant AYA2002-04094 and also supported  
by DURSI of the Generalitat de Catalunya. SEW acknowledges support from 
NASA (NNG05GG08G), and the DOE Program for Scientific Discovery through Advanced Computing (SciDAC; DE-FC02-01ER41176).





\clearpage

\begin{figure}
\plotone{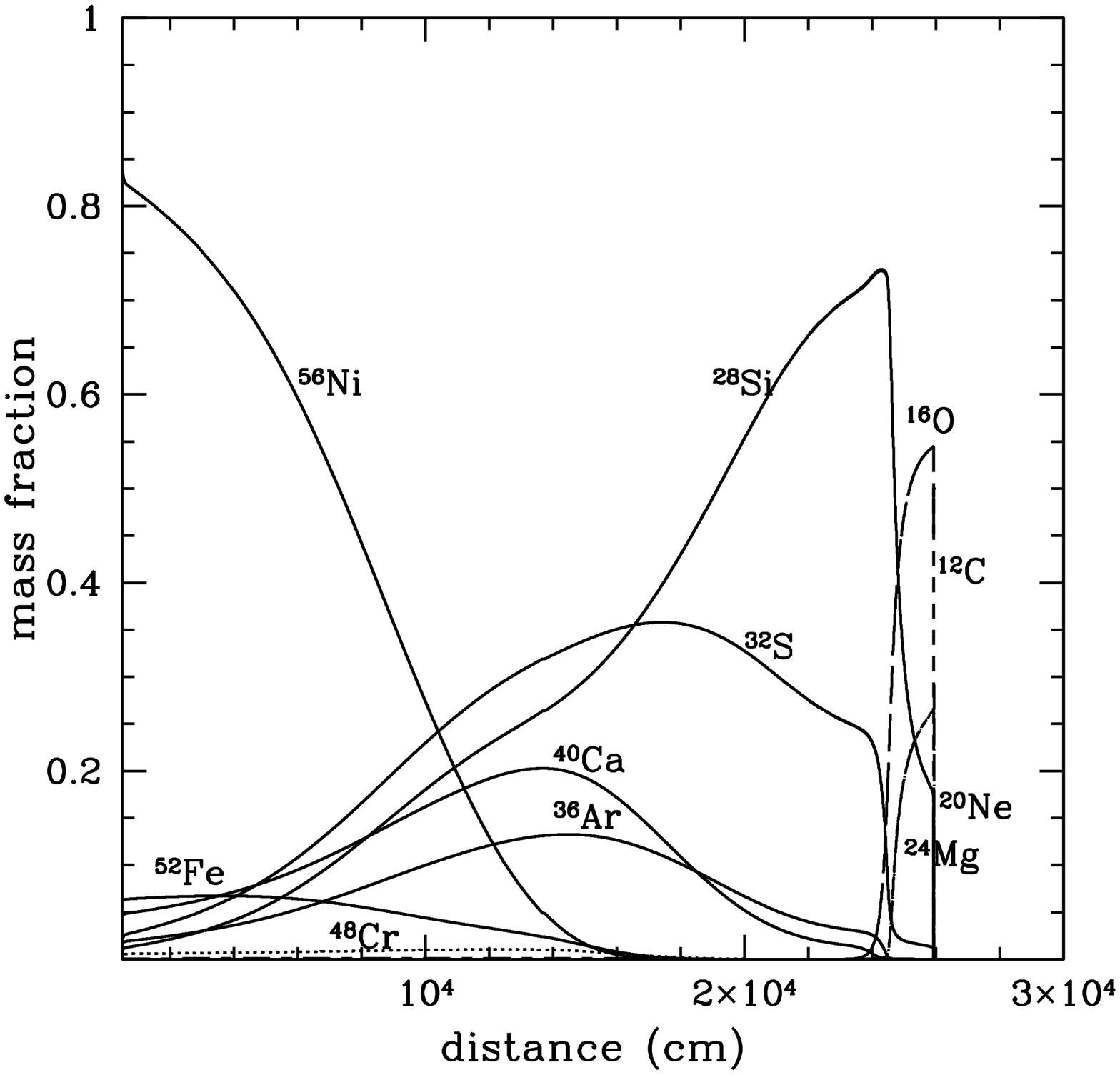}
\caption{Mass fraction profile of the most abundant species synthesized 
during the decay of a planar nuclear flame due to expansion. A 
characteristic expansion time $\tau_c=0.1 s$~was assumed in this case. The 
elapsed time was t=0.1523 s, which corresponds to a fuel density of
 $1.45\times 10^7$~g.cm$^{-3}$. Below that density  
the total amount of IME practically freezes}
\label{fig1}
\end{figure}

\clearpage
\begin{figure}
\plotone{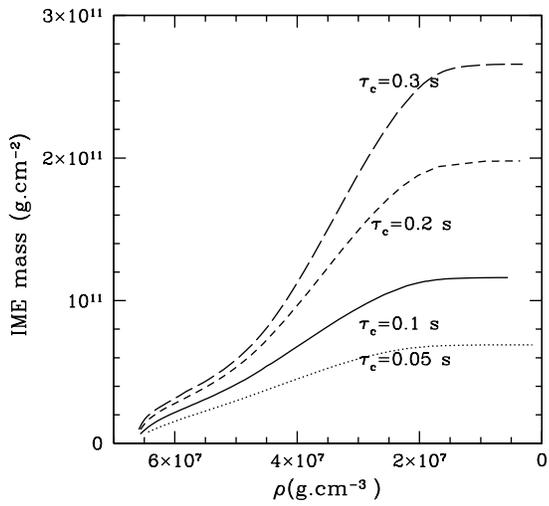}
\caption{Total mass per unit area of IME (from silicon to calcium) 
synthesized during 
the isobaric propagation of the flame as a function of the instantaneous 
density value  
in the cold fuel for four different characteristic expansion times. At $\rho=
2\times 10^7$~\dens practically all IME have already been produced. 
}
\label{fig2}
\end{figure}

\clearpage
\begin{figure}
\plotone{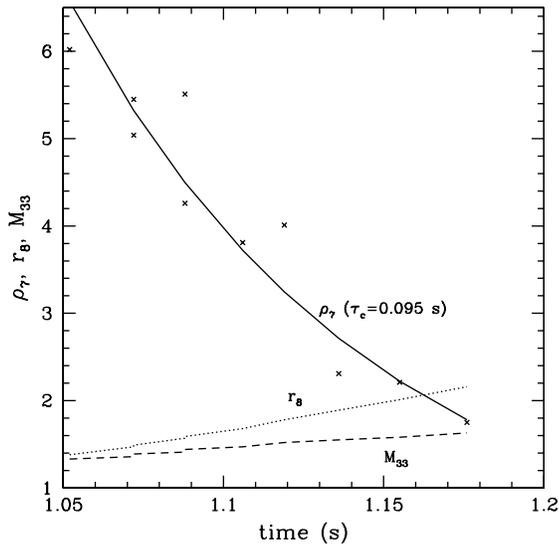}
\caption{Evolution of density, radius and lagrangian-mass coordinate (in $10^7, 
10^8 \mathrm {and}~ 10^{33}$~ c.g.s units respectively) of  
the one-dimensional model used to compute some of 
the inputs of Figure 4. The figure depicts the value of these variables 
just at the moment the temperature of the corresponding mass-shell becomes 
higher than one billion degrees. Thus, crosses represent the density of 
different mass-shells as they are reached by the flame. The solid line is an  
exponential fit to the numerical data using a 
characteristic time $\tau_c=0.095$~s. 
We have checked that the maximum 
temperature achieved by these shells is in the right range to give 
intermediate-mass elements.}  
\label{fig3}
\end{figure}

\clearpage
\begin{figure}
\plotone{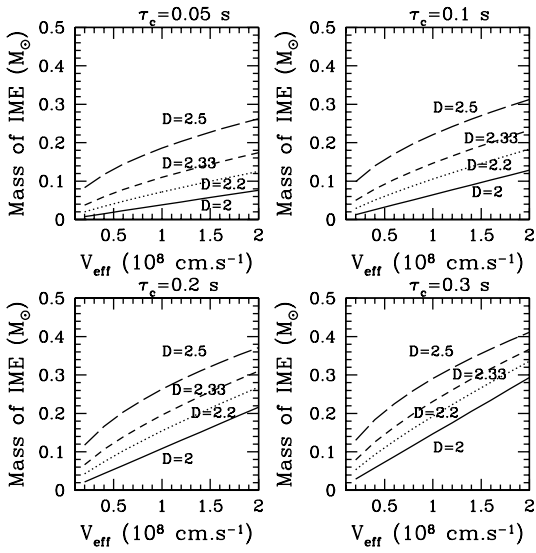}
\caption{Total amount (in M$_\odot$) of elements from silicon to calcium 
synthesized during the expansion of a massive white dwarf due to the 
propagation of a turbulent nuclear flame. The region    
where these nuclei are produced is approximated by a fractal of dimension 
D and average radius $\bar r=1.8\times 10^8$~cm. The quantity of IME is then 
given as a 
function of D, the effective propagation velocity of the flame $V_{\mathrm eff}$, and 
the characteristic expansion time $\tau_c$.
}  
\label{fig4}
\end{figure}

\clearpage
\begin{figure}
\plotone{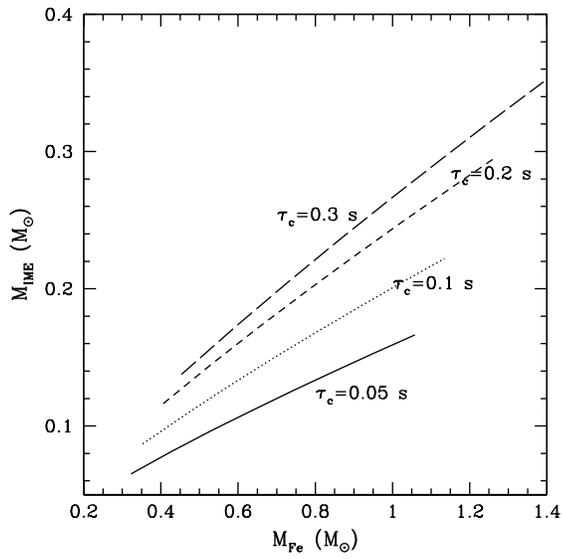}
\caption{Mass of IME as a function of the ejected mass of Fe-peak elements 
for several expansion times. A constant fractal dimension of the iron core
surface, 
D=7/3, was assumed as well as a constant turbulence 
intensity, V$=1.5\times 10^8$~cm.s$^{-1}$.
} 
\label{fig5}
\end{figure}

\clearpage
\begin{table*}
\caption{Final masses (g.cm$^{-2}$) of $\alpha$-elements produced\newline 
in a laminar expanding nuclear flame for \newline 
 several characteristic expansion times}  \label{Table 1}

\begin{tabular}{crrrrrr}
\hline
\noalign{\smallskip}
Nuclei&$\tau_c=0.05$~s&$\tau_c=0.1$~s&$\tau_c=0.2$~s&$\tau_c=0.3$~s\\  
& & -mass (g.cm$^{-2}$)- & & & \\
\hline
\noalign{\smallskip}
$^4$He &$4.1~10^{-2}$&$1.62~10^{-1}$&$4.25~10^5$&$2.7~10^7$\\
$^{12}$C &$3.3~10^{8}$&$3.48~10^{8}$&$3.72~10^8$&$3.9~10^8$\\
$^{16}$O &$4.9~10^{9}$&$7.7~10^{9}$&$1.23~10^{10}$&$1.6~10^{10}$\\
$^{20}$Ne &$3.5~10^{7}$&$4.4~10^{7}$&$5.60~10^7$&$6.9~10^7$\\
$^{24}$Mg &$1.9~10^{9}$&$3.2~10^{9}$&$5.12~10^9$&$6.9~10^9$\\
$^{28}$Si &$2.95~10^{10}$&$4.98~10^{10}$&$8.57~10^{10}$&$1.16~10^{11}$\\
$^{32}$S &$2.2~10^{10}$&$3.69~10^{10}$&$6.25~10^{10}$&$8.4~10^{10}$\\
$^{36}$Ar &$7.3~10^{9}$&$1.21~10^{10}$&$2.0~10^{10}$&$2.6~10^{10}$\\
$^{40}$Ca &$1.0~10^{10}$&$1.75~10^{10}$&$2.97~10^{10}$&$3.9~10^{10}$\\
$^{44}$Ti &$1.3~10^{7}$&$2.02~10^{7}$&$1.0~10^8$&$5.2~10^8$\\
$^{48}$Cr &$5.6~10^{8}$&$9.7~10^{8}$&$1.65~10^9$&$2.2~10^9$\\
$^{52}$Fe &$2.3~10^{9}$&$5.3~10^{9}$&$1.15~10^{10}$&$1.7~10^{10}$\\
$^{56}$Ni &$1.46~10^{10}$&$4.6~10^{10}$&$1.40~10^{11}$&$2.6~10^{11}$\\
$^{60}$Zn &$1.0~10^{3}$&$2.7~10^{3}$&$3.27~10^5$&$1.5~10^7$\\
\noalign{\smallskip}
\hline
\end{tabular}
\end{table*}

\clearpage
\begin{table*}
\caption{Estimated amounts of IME synthesized during an \newline   
exponential expansion from $\rho_0=6.6\times 10^7$~\dens \newline
 down to $\rho_0=2\times 10^6$~\dens 
with characteristic time $\tau_c$} 
 \label{Table 2}

\begin{tabular}{crrrrrrr}
\hline
\noalign{\smallskip}
$\tau_c$&$\Delta m_{burnt}$&$\Delta m_{IME}$&$\Delta m_{Ni}$~&X$_{IME}$&X$_{Ni}$&\\ 
 (s)& g.cm$^{-2}$&g.cm$^{-2}$&g.cm$^{-2}$& -- &--&\\
\noalign{\smallskip}
\hline
\noalign{\smallskip}
0.05 &$9.27~10^{10}$&$6.88~10^{10}$&$1.75~10^{10}$ &0.742 &0.188&\\
0.1 &$1.80~10^{11}$&$1.16~10^{11}$&$5.23~10^{10}$ &0.646&0.29&\\
0.2 &$3.69~10^{11}$&$1.98~10^{11}$&$1.53~10^{11}$ &0.537&0.415&\\
0.3 &$5.68~10^{11}$&$2.65~10^{11}$&$2.8~10^{11}$ &0.466&0.492&\\
\noalign{\smallskip}
\hline
\end{tabular}
\end{table*}


\clearpage
\begin{thebibliography}{}

\bibitem[Bravo et al. 1996]{bra96} Bravo, E., Tornamb\'e, A., Dominguez, I., \& Isern, J.   
 1996, A\&A, 306, 811.

\bibitem[2004]{cab04} Cabez\'on, R.M., Garc\'\i a-Senz, D.,  
 \& Bravo, E.  2005, ApJS, 151, 345.

\bibitem[Garc\'\i a-Senz \& Bravo 2005]{gar05} Garc\'\i a-Senz, D. \& Bravo, E.  
2005, A\&A, 430, 585

\bibitem[Iwamoto et al. 1999]{iwa99} Iwamoto, K., Brachwitz, F., Nomoto, K., Kishimoto, N., 
Umeda, H., et al. 1999, ApJS, 125, 439.

\bibitem[Khokhlov 1995]{kho95} Khokhlov, A.M. 1995, ApJ, 449, 695.

\bibitem[Khokhlov et al. 1997]{kho97} Khokhlov, A.M., Oran, E.S., \& Wheeler, J.C. 1997, ApJ, 478, 678

\bibitem[Niemeyer \& Woosley 1997]{nie97} Niemeyer, J.,\& Woosley, S.E. 1997, ApJ, 475, 740.

\bibitem[Niemeyer et al. 1999]{nie99} Niemeyer, J., Bushe, W.,K., \& Ruetsch, G.R. 1999, ApJ,
524, 290.

\bibitem[Nomoto et al. 1984]{nom84} Nomoto, K., Thielemann, F.-K., \& Yokoi, K.  1984, ApJ,
286, 644.

\bibitem[Peters \& Franke 1989]{pet89} Peters, N.,\& Franke, Ch. 1989, in Dissipative
 Structures, Transport Processes \& Combustion, ed. D. meink\"on (Berlin: 
Springer), 40.

\bibitem[Plewa et al. 2005]{ple04} Plewa, T., Calder, A.C., \& Lamb, D.Q. 2004, ApJ, 612, L37.
 
\bibitem[R\"opke et al. 2006]{rop06} R\"opke, F.K., Hillebrandt, W., Niemeyer, J.C., \& Woosley, S.E. 2006, A\&A, 448, 1.

\bibitem[Stritzinger et al. 2006]{str06} Stritzinger, M., Leibundgut, B., Walch,
S., \& Contardo, G. 2006, A\&A, 450, 241

\bibitem[1992]{tim92} Timmes, F.X.,\&  Woosley, S.E. 1992, ApJ, 396, 649.

\bibitem[Woosley 2001]{woo01} Woosley, S.E. 2001, Nucl. Phys., A688, 9c.

\bibitem[Zingale et al. 2005]{zin05} Zingale, M., Woosley, S.E., 
Rendleman, C.A., 
 Day, M.S., \& Bell, J.B. 2005, ApJ, 632, 1021.

\end{thebibliography}
\end{document}